\documentclass[aps,prl,twocolumn,showpacs,floatfix]{revtex4}
\usepackage{amsmath}
\usepackage{amsfonts}
\usepackage{amssymb}
\usepackage{epsfig}
\usepackage{graphicx}

\begin{document}

\title{Optomechanical Systems as Single Photon Routers}

\author{G. S. Agarwal and Sumei Huang}
\affiliation{Department of Physics, Oklahoma State University, Stillwater, Oklahoma 74078, USA}

\date{\today}

\begin{abstract}
We theoretically demonstrate the possibility of using nano mechanical systems as single photon routers. We show how electromagnetically induced transparency (EIT) in cavity optomechanical systems can be used to produce a switch for a probe field in a single photon Fock state using very low pumping powers of a few microwatts. We present estimates of vacuum and thermal noise and show that the optimal performance of the single photon switch is deteriorated by only a few percent even at temperatures of the order of 20 mK.
\end{abstract}


\maketitle
It is well known that the building of all optical devices requires strong interactions between radiation and matter as photons by themselves do not interact. One of the enabling technologies, in the context of quantum control, is the design of an optical switch or a photon router operating at a single photon level. In quantum information science one would like to distribute information over large distances. Photons are important candidates for such purposes as they can travel over long distances without much decoherence. Further, it is known that many quantum information protocols such as quantum cryptography~\cite{Bennett} and quantum networks~\cite{Kimble} require single photons. We need to route the photons, as different nodes in the networks have to be linked~\cite{Kimble}. Further the routing has to be done in a way so that the state of the photon is not affected. Several proposals have been made for the realization of an optical switch---In an early work Harris and Yamamoto \cite{Harris} had suggested how quantum interference can be used to operate a switch. More recently atomic EIT with cavity fields has been suggested to realize an optical switch. Single atom EIT in a cavity has been realized by using very strong atom cavity interactions \cite{Rempe}. Further, even vacuum induced transparency has been observed \cite{Vuletic}. Other proposals on a photon switch are based on using a single atom in a strongly coupled waveguide array \cite{Zhou} and using strongly coupled atoms via surface plasmons on a nanowire \cite{Lukin}. There are also reports of single photon switch at telecom wavelengths using strong cross phase modulation \cite{Kumar} and in the microwave domain using a superconducting transmon qubit \cite{Hoi}. It is now known that the optomechanical systems exhibit an analog of EIT \cite{Agarwal} which has been observed in several experiments \cite{Weis}. Herein, we show how nanomechanical mirrors (NMM) in optical cavities can be used to build single photon routers (i.e. single photon switches). For this purpose we use a configuration in which the NMM is in the middle of a cavity which is bounded by two high quality mirrors \cite{Thompson,add}. In a Fabry-Perot cavity a single photon will be transmitted if its frequency is on resonance with the cavity. Now we drive the cavity with a strong control field which is red detuned from the cavity resonance. In the presence of the strong control field, the NMM leads to reflection of the single photon. Thus the driving field switches the route of the single probe photon. This is in contrast to the situation where one uses a single photon to turn on and off the switch. Even a low driving field of a few microwatts is good. We present exact conditions for this to happen. We further investigate the effects of vacuum and thermal noise on the performance of this system as a single photon router. We show that the effect of these noise sources is only few percent at temperatures of the order of 20 mK. Although we concentrate on the routing of single photons, the analysis applies to other fields as well. Further we note that the single photon to be routed could be part of an entangled pair.

\begin{figure}[htp]
 \scalebox{0.6}{\includegraphics{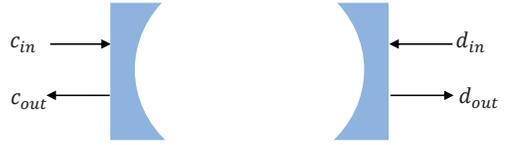}}
 \caption{\label{Fig1} (Color online) A double-ended cavity.}
 \end{figure}

Consider first a Fabry-Perot cavity where both mirrors have equal reflectivity. It is known that the transmission of a Fabry-Perot cavity goes to unity when the incident field is on resonance with the cavity. This result also applies if a single photon is incident on the cavity. In order to see this let us consider the input-output relations \cite{Walls} for the cavity as shown in Fig.~\ref{Fig1}.

\noindent Here $c_{in}$ and $d_{in}$ are the quantum fields incident on the cavity. If there are no photons incident from the right, then $d_{in}$ would be the vacuum field. Let $2\kappa$ be the rate at which photons leak out from each of the cavity mirrors. The out fields can be written as
\begin{equation}\label{1}
x_{out}(\omega)=\sqrt{2\kappa}c(\omega)-x_{in}(\omega),\ x=c\ \mbox{or}\ d,
\end{equation}
where the cavity field $c(\omega)$ is given by
\begin{equation}\label{2}
\dot{c}=-2\kappa c-i\omega_{0}c+\sqrt{2\kappa}(c_{in}+d_{in}).
\end{equation}
From (\ref{1}) and (\ref{2}) we find in steady state
\begin{equation}
c_{out}(\omega)=\frac{i(\omega-\omega_{0})c_{in}(\omega)+2\kappa d_{in}(\omega)}{2\kappa-i(\omega-\omega_{0})},\label{3}
\end{equation}
\begin{equation}
d_{out}(\omega)=\frac{2\kappa c_{in}(\omega)+i(\omega-\omega_{0})d_{in}(\omega)}{2\kappa-i(\omega-\omega_{0})}.\label{4}
\end{equation}
Here $d_{in}(\omega)$ is the vacuum field and hence its normally ordered correlation is zero. Defining the spectrum of the field via $\langle c^{\dag}(-\Omega)c(\omega)\rangle=2\pi S_{c}(\omega)\delta(\omega+\Omega)$, we obtain from (\ref{3}) and (\ref{4})
\begin{eqnarray}
S_{cout}(\omega)&=&\frac{(\omega-\omega_{0})^{2}}{4\kappa^{2}+(\omega-\omega_{0})^{2}}S_{cin}(\omega),\label{6}\\
S_{dout}(\omega)&=&\frac{4\kappa^{2}}{4\kappa^{2}+(\omega-\omega_{0})^{2}}S_{cin}(\omega).\label{7}
\end{eqnarray}
For $\omega=\omega_{0}$, $S_{cout}(\omega_{0}) \to 0$ and $S_{dout}(\omega_{0})=S_{cin}(\omega_{0})$. Therefore we have established that a single photon at the cavity's resonant frequency is completely transmitted. We next establish how a NMM in the cavity acts as a single photon router; i.e., by changing the intensity and frequency of the strong control laser, it would reflect the single probe photon $S_{cout}(\omega_{0})=S_{cin}(\omega_{0}),\ S_{dout}(\omega_{0})=0$.

\begin{figure}[htp]
 \scalebox{0.6}{\includegraphics{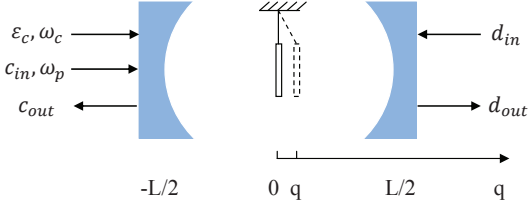}}
 \caption{\label{Fig2} (Color online) A double-ended cavity with a moving NMM as a single photon router.}
 \end{figure}
Consider now the configuration shown in Fig. \ref{Fig2} where a partially transparent NMM placed at the middle position of the Fabry-Perot cavity formed by two fixed mirrors which have finite identical transmission \cite{Thompson,add}. The whole cavity length is $L$. The cavity field is driven by a strong coupling field at frequency $\omega_{c}$ from the left-hand mirror. Further, a field in a single photon Fock state at frequency $\omega_{p}$ is incident on the cavity through the left-hand mirror. The input field is centered near the cavity frequency with a finite bandwidth $\Gamma$ i.e. its spectrum is given by $S_{cin}(\omega)=\frac{\Gamma/\pi}{(\omega-\omega_{0})^{2}+\Gamma^{2}}$.
The photons in the cavity will exert a radiation pressure force on the movable mirror, causing it to move. In turn, the displacement $q$ of the movable mirror shifts the cavity's resonance frequency. We assume that the movable mirror is located at the node of the cavity mode, thus the cavity's resonance frequency depends linearly on the displacement $q$ of the movable mirror. Here, the movable mirror is treated as a quantum harmonic oscillator with effective mass $m$, frequency $\omega_{m}$, and momentum operator $p$. Let $c$ and $c^{\dag}$ be the annihilation and creation operators for the cavity field. The Hamiltonian of the system in the rotating frame at the frequency $\omega_{c}$ of the coupling field is given by
\begin{equation}\label{11}
H=\hbar(\omega_{0}-\omega_{c})c^{\dag}c+\hbar g c^{\dag}c q+\frac{p^{2}}{2m}+\frac{1}{2}m\omega_{m}^{2}q^{2}+i\hbar\varepsilon_{c}(c^{\dag}-c),
\end{equation}
in which $g$ is the optomechanical coupling strength between the movable mirror and the cavity field, which also depends on the transmission of the movable mirror \cite{Thompson}. By choosing the transmission of the movable mirror $\mathcal{T}=0.7$, the optomechanical coupling strength can be half of that for a perfectly reflecting movable mirror so that $g=-\frac{\omega_{c}}{L}$ \cite{Meystre}. The driving strength $\varepsilon_{c}$, depends on the power $\wp$ of the coupling field, $\varepsilon_{c}=\sqrt{\frac{2\kappa\wp}{\hbar\omega_{c}}}$. Note that the movable mirror is coupled to the thermal surrounding at the temperature $T$, which results in the mechanical damping rate $\gamma_{m}$ and thermal noise force $\xi$ with frequency-domain correlation function:
\begin{equation}\label{12}
\langle \xi(\omega)\xi(\Omega)\rangle=2\pi\hbar\gamma_{m}m\omega\left[1+\coth\left(\frac{\hbar
\omega}{2k_{B}T}\right)\right]\delta(\omega+\Omega),
\end{equation}
where $k_B$ is the Boltzmann constant. In addition, the cavity field $c$ is coupled to the input quantum fields $c_{in}$ and $d_{in}$. These couplings are included in the standard way by writing quantum Langevin equations for the cavity field operators. The incoming vacuum field $d_{in}$ is characterized by $\langle d_{in}(\omega)
d_{in}^{\dag}(-\Omega)\rangle=2\pi \delta(\omega+\Omega)$ with $S_{din}(\omega)=0$. Putting together all the quantum fields, thermal fluctuations and the Heisenberg equations that follow from the Hamiltonian (\ref{11}), we obtain the working quantum Langevin equations
\begin{equation}\label{13}
\begin{aligned}
\dot{q}&=\frac{p}{m},\hspace{0.2in}\dot{p}=-\hbar g c^{\dag}c-m\omega_{m}^{2}q-\gamma_{m} p+\xi,\\
\dot{c}&=-[2\kappa+i(\omega_{0}-\omega_{c}+gq)]c+\varepsilon_{c}+\sqrt{2\kappa}c_{in}+\sqrt{2\kappa}d_{in},
\end{aligned}
\end{equation}
where mean values of noise terms $\xi$, $c_{in}$, and $d_{in}$ are zero.

Using the Langevin equations (\ref{13}) we want to calculate the spectrum of the output fields $c_{out}$ and $d_{out}$ so we adopt the standard quantum optical procedure \cite{Tombesi}. We first find the steady state for the mean values of the observable and then linearize the Langevin equations around the mean values to calculate quantum fluctuations. We quote the result of such a calculation and we find that the spectrum of the output fields has the form
\begin{eqnarray}\label{14}
S_{cout}(\omega)&=&S_{cin}(\omega)\cdot R(\omega)+S^{(v)}(\omega)+S^{(T)}(\omega),\nonumber\\
S_{dout}(\omega)&=&S_{cin}(\omega)\cdot T(\omega)+S^{(v)}(\omega)+S^{(T)}(\omega),
\end{eqnarray}
where
\begin{equation}\label{15}
R(\omega)=|E(\omega)-1|^{2},\hspace{0.2in} T(\omega)=|E(\omega)|^{2},
\end{equation}
and
\begin{eqnarray}\label{16}
S^{(v)}(\omega)&=&8\Big|\frac{\kappa}{d(\omega)}i\hbar g^{2}c_{s}^{2}\Big|^{2},\nonumber\\
S^{(T)}(\omega)&=&|V(\omega)|^{2}\hbar\gamma_{m}m(-\omega)\left[1+\coth\left(-\frac{\hbar\omega}{2k_{B}T}\right)\right],\nonumber\\
E(\omega)&=&\frac{2\kappa}{d(\omega)}\{m(\omega_{m}^{2}-\omega^{2}-i\gamma_{m}\omega)[2\kappa-i(\Delta+\omega)]\nonumber\\& &+i\hbar g^{2}|c_{s}|^{2}\},\nonumber\\
V(\omega)&=&\frac{\sqrt{2\kappa}}{d(\omega)}\{-igc_{s}[2\kappa-i(\Delta+\omega)]\},\nonumber\\
d(\omega)&=&m(\omega_{m}^{2}-\omega^{2}-i\gamma_{m}\omega)[(2\kappa-i\omega)^{2}+\Delta^{2}]\nonumber\\& &-2\hbar g^{2}|c_{s}|^{2}\Delta,
\end{eqnarray}
where $q_{s}=-\frac{\hbar g|c_{s}|^{2}}{m\omega_{m}^{2}}$, $c_{s}=\frac{\varepsilon_{c}}{2\kappa+i\Delta}$, and $\Delta=\omega_{0}-\omega_{c}+gq_{s}$ is the effective detuning which includes the frequency shift due to radiation pressure. $|c_{s}|^{2}$ is the number of intracavity photons, and $q_{s}$ is the steady state position of the movable mirror. The roots of $d(\omega)$ determine essentially the behavior of the output fields; they are complex and depend on the power of the coupling field.

In Eq. (\ref{14}), $R(\omega)$ and $T(\omega)$ are the contributions arising from the presence of a single photon in the input field. The $S^{(v)}(\omega)$ is the contribution from the incoming vacuum field. The nonlinear coupling of the cavity field with the mirror converts the vacuum
photon at frequency $\omega_{c}-\Omega$ to $\omega_{c}+\Omega$ via the mixing process $\omega_{c}+\omega_{c}-(\omega_{c}-\Omega)\to\omega_{c}+\Omega$. Note that $S^{(v)}(\omega)$ is at least of order four in the amplitude of the coupling field and this determines the nature of the vacuum contribution. The $S^{(T)}(\omega)$ is the contribution from the fluctuations of the mirror. The equation (\ref{14}) shows that even if there were no incoming photon, the output signal is generated via quantum and thermal noises. For the purpose of achieving single photon router, the key quantities are $R(\omega)$ and $T(\omega)$. Further, the performance of the single photon router should not be deteriorated by the quantum and thermal noise terms $S^{(v)}(\omega)$ and $S^{(T)}(\omega)$. We also note in passing that in a treatment where the probe field is treated classically, the output semiclassical fields would be $(E(\omega)-1)$ and $E(\omega)$ on the left and right ports, respectively.

\begin{figure}[htp]
 \scalebox{0.6}{\includegraphics{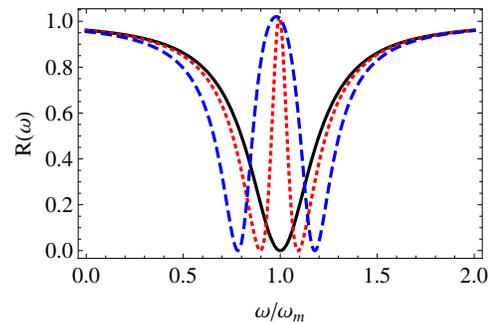}}
 \caption{\label{Fig3} (Color online) The reflection spectrum $R(\omega)$ of the single photon as a function of the normalized frequency $\omega/\omega_{m}$ without and with the coupling field. $\wp=0$ (solid), 5 $\mu$W (dotted), 20 $\mu$W (dashed).}
 \end{figure}

\begin{figure}[htp]
 \scalebox{0.6}{\includegraphics{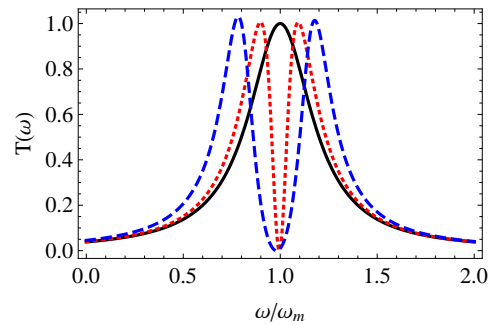}}
 \caption{\label{Fig4} (Color online) The transmission spectrum $T(\omega)$ for the same parameters as in Fig.~\ref{Fig3}.}
 \end{figure}

Using the results (\ref{14}), we now present numerical results for the spectra of the transmitted and reflected single photon field. We work in the sideband resolved limit (i.e. $\omega_{m}\gg\kappa$) and further we will take $\Delta=\omega_{m}$. We use the parameters \cite{Thompson}. The wavelength of the coupling field $\lambda=1054$ nm, $L=6.7$ cm, $m=40$ ng, $\omega_{m}=2\pi\times 134$ kHz, $Q=1.1\times 10^{6}$, $\gamma_{m}=\omega_{m}/Q=0.76$ sec$^{-1}$, and $\kappa=\omega_{m}/10$. In the following, we work in the stable regime of the system i.e. we use control power such that there are no instabilities. The resulting spectra are shown in Figs.~\ref{Fig3} and \ref{Fig4}.
In the absence of the coupling field, one observes an inverted Lorentzian and a standard Lorentzian in the reflection and transmission spectra of a single photon. Note that $R(\omega_{m})\approx0$ and $T(\omega_{m})\approx1$. So the single photon is completely transmitted through the cavity to the right output port. However, in the presence of the coupling field, the situation is completely different. The reflection and transmission spectra of a single photon exhibit an inverted dip and a dip at $\omega=\omega_{m}$, and $R(\omega_{m})\approx1$ and $T(\omega_{m})\approx0$. The single photon is totally reflected to the left output port. In the presence of the coupling field, the NMM participates in the transmission or reflection of the photon and we have all the conditions for occurrence of EIT fulfilled $(\gamma_{m}\ll \kappa\ll\omega_{m},\ \omega_{p}=\omega_{c}+\omega_{m})$. Therefore the incident single photon is totally reflected. In an earlier work~\cite{Agarwal}, we dealt with coherent light and it was shown that the reflected outgoing field would even have a well defined phase; however, for router action, phase does not play a role. Note further that in the results (\ref{14}), the incident quantum field appears via the quantity $S_{cin}(\omega)$. Thus in principle one could also treat the switching action of other quantum fields as long as their spectral width falls within the EIT dip. The final outcome would depend on the overlap of the input spectrum with $R(\omega)$ and $T(\omega)$. Our calculation shows that the width of the transmission dip or reflection peak is $\frac{\gamma_{m}}{2}+\frac{\hbar g^{2}|\varepsilon_{c}|^{2}}{4m\omega_{m}\kappa(4\kappa^{2}+\omega_{m}^{2})}$, which for the parameters of Fig.~\ref{Fig3} is about 0.05 $\omega_{m}$ and 0.23 $\omega_{m}$ for $\wp=5$ $\mu$W and 20 $\mu$W, respectively.
\begin{figure}[htp]
 \scalebox{0.5}{\includegraphics{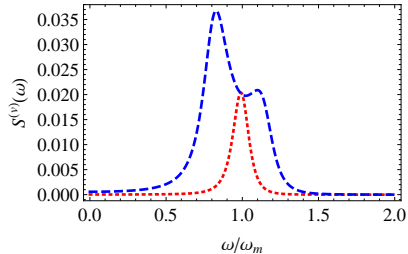}}
 \caption{\label{Fig5} (Color online) The vacuum noise spectrum $S^{(v)}(\omega)$ as a function of the normalized frequency $\omega/\omega_{m}$ with the coupling field. $\wp=5$ $\mu$W (dotted), 20 $\mu$W (dashed).}
 \end{figure}

\begin{figure}[htp]
 \scalebox{0.5}{\includegraphics{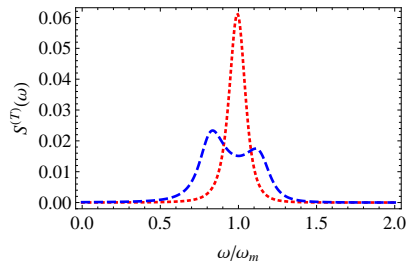}}
 \caption{\label{Fig6} (Color online) The thermal noise spectrum $S^{(T)}(\omega)$. The other parameters are the same as in Fig.~\ref{Fig5}.}
 \end{figure}

Next we discuss the effects of the quantum and thermal noises on the reflection and transmission spectra of a single photon. We exhibit the behavior of the vacuum noise $S^{(v)}(\omega)$ for two different values of the coupling power in Fig.~\ref{Fig5}. The contribution of the vacuum noise is about 2 \% at $\omega=\omega_{m}$ and is thus insignificant. Note that for larger coupling powers, $S^{(v)}(\omega)$ splits into two peaks---this is connected with the normal mode splitting \cite{Aspelmeyer}, arising from the two roots of $d(\omega)$. The thermal noise could be more critical in deteriorating the performance of the single photon router. Clearly to beat the effects of thermal noise, the number of photons in the probe pulse has to be much bigger than the thermal noise photons. However if we work with mirror temperatures like 20 mK, then the thermal noise term is insignificant as shown in Fig. \ref{Fig6}. It is worth noting that higher powers or equivalently cavities with larger coupling constant are better for the suppression of thermal noise. Even at a relatively large temperature, for example 200 mK, the maximum thermal noise is about 20 \% at $\wp=20$ $\mu$W. In the light of rather small sources of noise at temperatures $\lesssim$ 100 mK, we conclude that the NMM in an optical cavity as a single photon router is an excellent device.

In conclusion, we have shown how a cavity optomechanical system can be used as a single photon router~\cite{add2}. The physical mechanism that enables this application is the EIT behavior that this system exhibits. Further, we have shown that the effects of quantum noise sources on such a single photon router are very minimal.

GSA thanks S. Dutta Gupta and Anjali Agarwal for the discussions on single photon switch, and the Director of the Tata Institute of Fundamental Research, Mumbai for the hospitality where part of this work was done.


\begin{thebibliography}{99}
\bibitem{add} We choose three mirror geometry as we need two distinct paths for photons as in many of the two mirror geometries~\cite{Aspelmeyer} the nanomirror is highly reflecting, though the physics is similar.
\bibitem{add2} An interesting question concerns switching time of our system. This requires considerations of time dependent control fields. Our numerical simulations show that the switching from transmission to reflection follows the switching of the control pulse.
\end{thebibliography}
\end{document}